\documentclass[aps,prl,twocolumn,groupedaddress,letterpaper]{revtex4}
\usepackage[pdftex]{graphicx}
\usepackage{latexsym}
\begin{document}

\title{ Modifying Fragility and Collective Motion in Polymer Melts with Nanoparticles} 
\author{Francis W. Starr}
\affiliation{Department of Physics, Wesleyan University, Middletown, CT
  06459, USA} \author{Jack F. Douglas}
%\thanks{Official contribution of the National Institute of Standards and
%    Technology - Not subject to copyright in the United States.}
\affiliation{Polymers Division, National Institute of Standards and
  Technology, Gaithersburg, Maryland 20899, USA}

\date{Submitted 31 August 2010}
 
\begin{abstract}
  
  We investigate the impact of nanoparticles (NP) on the fragility and
  cooperative string-like motion in a model glass-forming polymer melt
  by molecular dynamics simulation.  The NP cause significant changes to
  both the fragility and the average length of string-like motion, where
  the effect depends on the NP-polymer interaction and the NP
  concentration. We interpret these changes via the Adam-Gibbs (AG)
  theory, assuming the strings can be identified with the
  ``cooperatively rearranging regions'' of AG.  Our findings indicate
  fragility is primarily a measure of the temperature dependence of the
  cooperativity of molecular motion.

\end{abstract}

\maketitle

%\begin{widetext}
% put single column material here;  iffy tool
%\end{widetext}

The addition of a small concentration of nanoparticles (NP) to
glass-forming (GF) polymer materials can lead to large property changes
that are difficult to comprehend by extension of the effects of
macroscopic filler additives. Depending on system details, changes may
be rationalized by the large surface-to-volume ratio of NP, chain
bridging~\cite{schweizer,gersappe,kds07}, or NP self-assembly into
extended structures~\cite{sdg03,akcora09}.  Changes in the glass
transition temperature $T_g$ have been emphasized, and both experimental
and theoretical studies indicate that attractive or repulsive
(non-attractive) polymer-NP interactions tend to increase or decrease
$T_g$, respectively.  Correspondingly, the interfacial polymer layer
around the NP shows a slowing down (increased $T_g$) or an acceleration
of dynamics (decreased $T_g$), providing a molecular
interpretation~\cite{bansal05,fd01,ssg,tnp00}.

Unfortunately, $T_g$ changes provide only a limited understanding of how
NP affect the properties of GF polymer melts. We also expect that the
$T$ dependence of dynamical properties approaching $T_g$, or the
``fragility'' of glass formation~\cite{angell95}, will be altered.
Fragility changes have been argued for on theoretical
grounds~\cite{dfd}, based on the finding that changes in the molecular
packing in the glass state ($T < T_g$) should also alter the fragility
of glass formation.  The NP we study should be particularly effective at
modifying molecular scale packing and motion, since their size is
roughly commensurate with the heterogeneity scale of fluids near
$T_g$~\cite{dyn-het}.
%(i.e., 2~nm to 3~nm)

In this letter, we address how polymer-NP interactions and NP
concentration affect the fragility of glass formation, and how this
effect relates to cooperative motion.  We demonstrate that the changes
in the relaxation time $\tau$ can be related to changes in the average
size $L$ of the string-like cooperative motion of monomers.  The
Adam-Gibbs (AG) theory~\cite{ag}, predicts a specific relationship
between between size of hypothetical ``cooperatively rearranging
regions'' (CRR) and structural relaxation time.  If $L$ corresponds the
size of the CRR, we find that the AG relation holds for all
concentrations $\phi$ and interaction types considered.  Additionally,
the AG theory predicts that fragility is sensitive to the $T$ dependence
of the size of CRR, and we confirm this relationship.

Our findings are based on equilibrium molecular dynamics simulations of
a nanoparticle surrounded by a dense polymer melt, as well as
simulations of a pure melt for comparison purposes. We utilize periodic
boundary conditions so that our results represent an ideal, uniform
dispersion of NP.  The polymers are modeled by a well-studied
bead-spring model~\cite{fene2}, but with the cut-off distance between
pairs extended to include attractive Lennard-Jones (LJ) interactions.
All monomer pairs interact via a LJ potential, and bonded monomers along
a chain are connected via a FENE anharmonic spring potential.  The NP
consists of 356 Lennard-Jones particles bonded to form an icosahedral
NP; the facet size of the NP roughly equals the equilibrium end-to-end
distance for a chain of 20 monomers.  Details of the simulation protocol
and our model potentials can be found in the supplemental material and
in ref.~\cite{ssg}.

We simulate systems with 100, 200, or 400 chains of $M=20$ monomers each
(for totals of $N=2000$, 4000, and 8000 monomers) to address the effect
of varying the NP volume fraction.  Under constant pressure conditions,
the addition of nanoparticles can give rise to a change in the overall
melt density.  A slight change in density can cause a significant change
in the dynamic properties relative to the pure melt.  In order to probe
only changes caused by the interactions between the NP and the polymer
melt, we have matched the density of monomers far from the NP with that
of the pure polymer melt~\cite{ssg}.

\begin{figure}[t]
\begin{center}
\includegraphics[clip,width=3.3in]{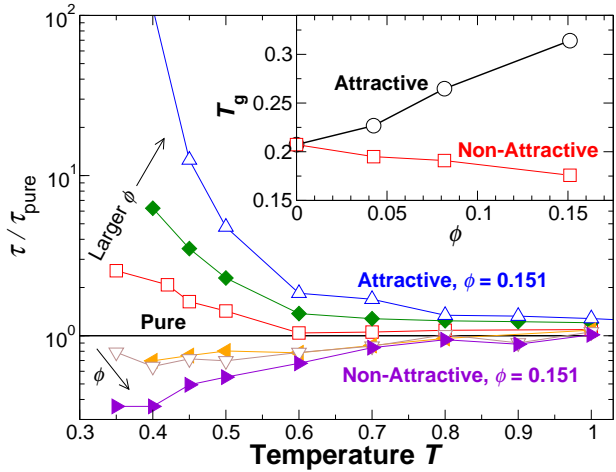}
\end{center}
\caption{Structural relaxation time $\tau$ as a function of $T$ for each
  $\phi$ normalized by $\tau_{\rm pure}$ for the pure melt. The inset
  shows the corresponding $T_g$ as a function of concentration $\phi$.
  Attractive interactions increase $\tau$ and $T_g$, while
  non-attractive interactions decrease $\tau$ and $T_g$.  In both cases,
  the effect is more pronounced with increasing concentration.  The
  concentrations are $\phi = 0.0426$ (red $\bigcirc$ and orange
  $\bigtriangleup$), 0.0817 (green $\Box$ and brown $\triangleleft$) and
  0.151 (blue $\Diamond$ and violet $\bigtriangledown$).  The pure melt
  is indicated in black.  The fits of the VFT relation to the data
  deviate by at most 0.5~\%.}
\label{fig:tau}
\end{figure}

To quantify changes in the nanocomposite dynamics, we evaluate the
effect of $\phi$ and the polymer-NP interactions on $\tau$, measured
from the relaxation of the coherent intermediate scattering function
(see supplementary information).  The effects of interactions on $\tau$
and $T_g$ for some $\phi$ were presented in ref.~\cite{ssg}; here we
provide additional simulation data and focus our analysis on fragility
and cooperative motion.  As expected, Fig.~\ref{fig:tau} shows that
attractive polymer-NP interactions slow the relaxation ($\tau$ becomes
larger), while non-attractive polymer-NP interactions give rise to an
increased rate of relaxation ($\tau$ becomes smaller).  The effect of
$\phi$ is more clearly seen by rescaling $\tau$ by the value $\tau_{\rm
  pure}$ in the pure melt, which shows that $\tau$ can be altered by a
factor of more than an order of magnitude on cooling.  The effect of the
NP is more pronounced at low $T$.

We next examine how these changes in $\tau$ affect $T_g$ and fragility.
For reference, the inset of Fig.~\ref{fig:tau} confirms that $T_g$
increases when there is attraction, and decreases with non-attractive
interactions. To estimate $T_g$, we fit the data using the
Vogel-Fulcher-Tammann (VFT) expression~\cite{angell95}
$\tau \propto  \exp{(D/[T/T_0-1])}.$
$T_0$ is an extrapolated divergence $T$ of $\tau$, while $D$
provides one measure of the fragility.  
We use the VFT fit to estimate $T_g$ based on the condition that
$\tau(T_g) = 100$~s (the canonical definition of the laboratory glass
transition~\cite{angell95}), assuming the one time unit in standard LJ
reduced units corresponds to 1~ps (reduced units defined in
supplementary information).

\begin{figure}[t]
\begin{center}
\includegraphics[clip,width=3.3in]{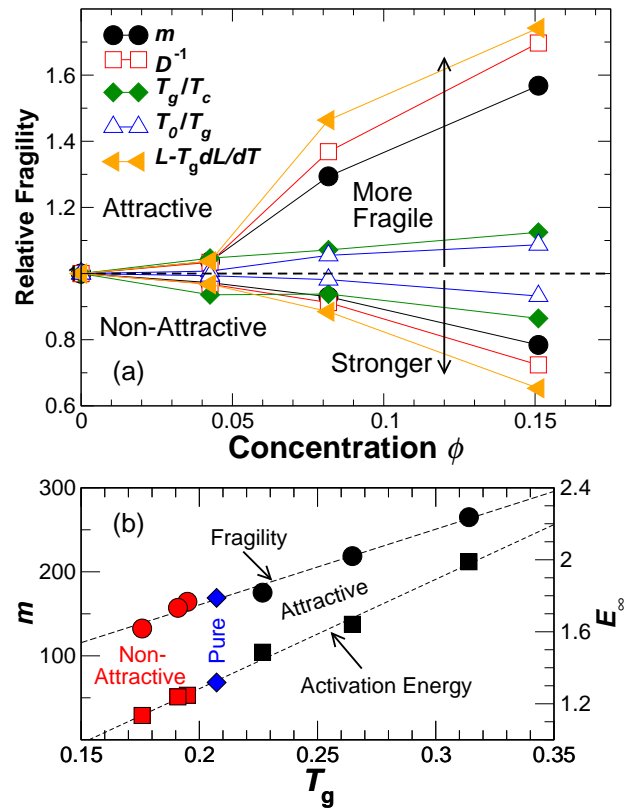}
\end{center}
\caption{(a) Fragility dependence on concentration $\phi$ relative to
  the pure melt.  We consider five different measures of fragility,
  which are discussed in the text.  All measures of fragility show the
  same qualitative trend: namely, the system with attractive polymer-NP
  interactions becomes more fragile, while the system with
  non-attractive interactions becomes less fragile (stronger).  The last
  measure, related to the string length $L$, is discussed later in the
  text.  (b) Demonstration of the proportionality between $T_g$ and
  fragility $m$, as well as the high-$T$ activation energy $E_\infty$.
  The different colors of symbols indicate the nature of the polymer-NP
  interactions. }
\label{fig:fragility}
\end{figure}

Since there is no single agreed upon measure of fragility, we consider
several different measures to ensure consistency. First, as indicated
above, the parameter $D$ from a VFT fit to $\tau$ is widely utilized;
specifically, a larger value of $D$ indicates a stronger (less fragile)
GF fluid so that $D^{-1}$ increases with increasing fragility. 
The most common definition of fragility is based on the $T$
dependence of $\tau$ near $T_g$, namely~\cite{angell85},
\begin{equation}
m=\left.\frac{d(\ln\tau)}{d(T_g/T)}\right|_{T_g}.
\label{eq:fragility}
\end{equation}
For strong GF systems, the rate of change of $\tau$ with respect to $T$ is
smaller than that of fragile systems; hence $m$ is larger for more
fragile GF fluids.  We estimate $m$ using our VFT fit.  Fragility can
also be estimated by the ratios $T_0/T_g$ or $T_g/T_c$.
We estimate $T_c$ using the power-law form 
%\begin{equation}
$\tau \sim (T/T_c-1)^{-\gamma}$
%\end{equation}
in an appropriate $T$ range (see supplementary information for fitting
details).
$T_0/T_g$ and $T_g/T_c$ are larger in more fragile systems~\cite{dfd}.

We summarize the results for the various fragility metrics in
Fig.~\ref{fig:fragility}(a), where we find that -- for all definitions
-- attractive polymer-NP interactions lead to more fragile glass
formation as a function of $\phi$; conversely, non-attractive polymer-NP
interactions lead to stronger glass formation.  These changes in
fragility mirror the changes in $T_g$ (fig.~\ref{fig:tau} inset).  In
particular, fig.~\ref{fig:fragility}(b) shows that $m \propto T_g$,
consistent with experimental trends in pure polymeric glass
formers~\cite{mckenna} and analytic calculations based on the entropy
theory of glass formation~\cite{stukalin}.
Figure~\ref{fig:fragility}(b) also shows that the high-$T$ activation
energy $E_\infty$ is roughly proportional to $T_g$.  We discuss the
implications of these scaling relationships below.

Our findings for fragility changes are consistent with experimental
studies of polymer-NP systems.  Bansal et al.~\cite{bansal05} found that
dispersions of NP having repulsive interactions caused $T_g$ to
decrease, accompanied by an appreciable broadening of the glass
transition region, indicative of increased strength (decreased
fragility) of glass formation.
For fullerenes dispersed in polystyrene, Cabral and
co-workers~\cite{cabral} reported behavior expected for attractive
polymer-NP interactions, namely an increase in $T_g$, accompanied by an
increased fragility.  For small $\phi$, negligible changes in the
fragility have been reported~\cite{green}, also consistent with our
small $\phi$ results.  Our results are likely not applicable when the
NP-polymer interactions are so strong that non-equilibrium effects
(leading to ``bound'' polymer)~\cite{harton} or phase separation may
dominate.

We next examine how the NP interactions impact the heterogeneity of
molecular motions and how this relates to the observed changes in $T_g$
and fragility.
Both small-molecule and polymeric liquids exhibit pronounced spatial
correlations in mobility, commonly referred to as ``dynamical
heterogeneity''~\cite{dyn-het}.  In particular, the most mobile atoms or
molecules
tend the cluster on a time scale after the ``breaking of the cage'', but
before the primary relaxation $\tau$.  These clusters of mobile
molecules can be further dissected into ``strings'' involving particles
moving roughly co-linearly~\cite{strings}; these structures appear to be
the most basic units of cooperative relaxation.  The characteristic size
of both the mobile-particle clusters and strings grow as a fluid is
cooled toward $T_g$. Notably, the string-like collective motion is not
strongly correlated with chain connectivity~\cite{strings-polymer}, so
it should not be confused with reptative motion.

\begin{figure}[t]
\begin{center}
\includegraphics[clip,width=3.3in]{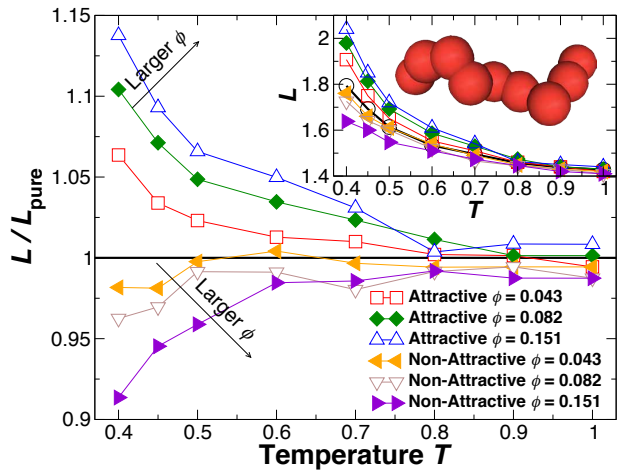}
\end{center}
\caption{$T$ dependence of $L$ for different $\phi$ normalized by $L$
  for the pure melt. Note the parallelism to fig.~\ref{fig:tau}.  The
  inset also shows a snapshot of an example string of length $L=9$.}
\label{fig:L-norm}
\end{figure}

We evaluate the string size $L(T)$ following the procedures developed in
ref.~\cite{strings}, which are slightly modified for the specific case
of the bead-spring polymer we study, as discussed in
ref.~\cite{strings-polymer} (see supplementary material for the
technical details).  
Figure~\ref{fig:L-norm} shows $L$ for all $T$ and $\phi$, where we
compare systems by normalizing by the value of $L$ of the pure
melt. Given that $L$ grows as $\tau$ grows on cooling, we expect that
the attractive NP interactions (which increase $\tau$ relative to the
pure melt) should cause $L$ to increase relative to the pure melt, and
vice-versa for non-attractive NP interactions.  Indeed, the variation of
$L$ is consistent with these expectations, and we conclude that the the
attractive NP interactions cause an increase in the degree of correlated
molecular motion for fixed $T$, while non-attractive NP interactions
cause a decrease in $L$. Note that the $T$ dependence of $L$ is
qualitatively similar in all cases.

While the changes in $L$ reflect the changes in $\tau$ for any given
$T$, how do the $T$ dependence of these changes compare?  In other
words, can $L$ be used to {\it predict} the fragility changes in
fig.~\ref{fig:fragility}(a)?  To answer this question, we are guided by
AG theory, which proposes that relaxation in GF liquids is dominated by
cooperatively rearranging regions (CRR). Specifically, AG argue that
$\tau$ is related to the average number of particles in the CRR $z$ by a
generalized Arrhenius relation~\cite{AG-footnote},
\begin{equation}
\tau = \tau_\infty \exp(z E_\infty/T).
\label{eq:AG-strings}
\end{equation}
For high $T$, cooperativity is minimal, so $z \approx 1$, and $E_\infty$
(a constant) can be identified with the high-$T$ activation energy of
the fluid.  Accordingly, the $T$-dependent activation energy $E(T) =
z(T) \, E_\infty$ should govern the fragility.  

\begin{figure}[t]
\begin{center}
\includegraphics[clip,width=3.3in]{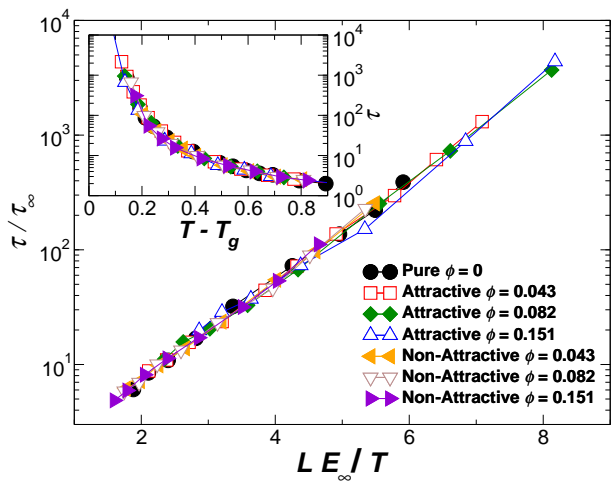}
\end{center}
\caption{Test of the AG relation (eq.~(\ref{eq:AG-strings})) between the
  size of CRR and the structural relaxation time $\tau$, assuming
  $z\propto L$.  The main figure shows the data scaled by the fit
  parameters so that all data sets collapse to a master curve. $\tau_\infty$
  is defined by eq.~\ref{eq:AG-strings}, replacing $z$ by $L$.
  The inset shows that $\tau$ can be collapsed by plotting as a function
  of $T-T_g$, a result of the proportionality of $T_g$ and $m$ (see
  fig.~\ref{fig:fragility}(b)). }
\label{fig:L-tau}
\end{figure}

The strings are a natural candidate to describe the abstract CRR of AG.
Fig.~\ref{fig:L-tau} shows that eq.~(\ref{eq:AG-strings}) with $z
\propto L$ provides an excellent prediction for $\tau$, consistent with
identifying $L$ with the CRR of AG~\cite{gbss03}.  Therefore, the
non-Arrhenius behavior of $\tau$ can be viewed as a consequence of the
increase in $L$ on cooling.  Accordingly, $L$ must encode the fragility
of glass formation.  In particular, combining eqs.~(\ref{eq:fragility})
and (\ref{eq:AG-strings}) implies a direct relation between $m$ and $L$,
\begin{equation} 
m = (E_\infty/T_g) \, \left[L(T_g)  - T_g\, dL/dT|_{T_g}\right].
\label{eq:AG-fragility}
\end{equation}
Since $E_\infty$ and $T_g$ are proportional
(fig.~\ref{fig:fragility}(b)), the prefactor $E_\infty/T_g$ is roughly
constant for all $\phi$, and thus does not impact the nanocomposite
fragility relative to the pure melt.  Therefore, the fragility changes
measured by $m$ on adding NP should result from changes to $L$ and $T \,
dL/dT$ at $T_g$.  To test this, we extrapolate $L$ to $T_g$ by assuming
consistency between eq.~(\ref{eq:AG-strings}) and the VFT expression.
Figure~\ref{fig:fragility}(a) shows that $L - T_g dL/dT$ indeed accounts
well for the observed changes in $m$.  In particular, the contribution
from $T_g\, dL/dT$ is 3 to 8 times larger than $L$, which ranges from
4.5 to 6.5 at $T_g$.  The range for $L$ is quantitatively consistent
with ref.~\cite{dfd}, and qualitatively consistent with experimental
evidence for weak sensitivity of fragility to the scale of
cooperativity~\cite{berthier-science,sokolov-coop}.  Hence, near $T_g$,
fragility is primarily controlled by $dL/dT$, rather than $L$,
consistent with ref.~\cite{dfd06}.  Stated more plainly, fragility is
primarily a measure of the rate of change $dL/dT$ of the extent of
cooperative motion.

The approximate proportionality between $m$ and $T_g$ observed in in our
system (Fig.~\ref{fig:fragility}(b)), and for many high molecular mass
polymer polymer materials~\cite{mckenna}, implies an important
simplification.  Specifically, since $m \sim 1/ D$, the product $DT_g$
should be nearly constant, and hence $\tau$ should be a nearly universal
function of $(T -T_g)$. The inset to Fig.~\ref{fig:L-tau} shows that
such a shift collapses all our nanocomposite data rather well; this
alternate data reduction also indicates the consistency of our $T_g$
extrapolation.  However, we are careful to point out that the
proportionality $m\propto T_g$ is not
universal~\cite{sokolov-fragility}, and this trend can even be reversed
in polyelectrolyte materials of interest in battery
applications~\cite{runt,stukalin}.

In summary, the addition of NP to a polymer melt can lead to significant
changes in both $T_g$ and fragility, which can be related to the
cooperative string-like motion.  Our results support the identification
of the strings with the abstract CRR of the AG theory, and complement
tests of the entropy formulation of the AG theory~\cite{AGpapers}.


\begin{thebibliography}{1}
\bibitem{AG-footnote} The original AG arguments do not discriminate
  between relaxation times associated with the mass, momentum and energy
  diffusion processes.  Since AG emphasize activated mass transport, it
  is natural to apply their arguments to a mass diffusion relaxation
  time. However, the decoupling relation between structural
  $\alpha$-relaxation $\tau$ and diffusivity implies that AG can also be
  applied to $\tau$, but with a modified high temperature activation
  energy.
\end{thebibliography}
\end{document}